# Excitation-Dependent High-Lying Excitonic Exchange *via* Interlayer Energy Transfer from *Lower-to-Higher* Bandgap 2D Material

*Arka Karmakar*[1*], *Tomasz Kazimierczuk*[1], *Igor Antoniazzi*[1], *Mateusz Raczyński*[1], *Suji Park*[2], *Houk Jang*[2], *Takashi Taniguchi*[3], *Kenji Watanabe*[4], *Adam Babiński*[1], *Abdullah Al-Mahboob*[2,↓], *Maciej R. Molas*[1#]

[1] Division of Solid State Physics, Institute of Experimental Physics, Faculty of Physics, University of Warsaw, Pasteura 5, 02-093 Warsaw, Poland

[2] Center for Functional Nanomaterials, Brookhaven National Laboratory, Upton, NY 11973, USA

[3] International Center for Materials Nanoarchitectonics, National Institute for Materials Science, 1-1 Namiki, Tsukuba, Ibaraki 305-0044, Japan

[4] Research Center for Functional Materials, National Institute for Materials Science, 1-1 Namiki, Tsukuba, Ibaraki 305-0044, Japan

[*] arka.karmakar@fuw.edu.pl

[↓] aalmahboo@bnl.gov

[#] maciej.molas@fuw.edu.pl

High light absorption (~15%) and strong photoluminescence (PL) emission in monolayer (1L) transition-metal dichalcogenide (TMD) make it an ideal candidate for optoelectronic applications. Competing interlayer charge (CT) and energy transfer (ET) processes control the photocarrier relaxation pathways in TMD heterostructures (HSs). In TMDs, long-distance ET can survive up to several tens of nm, unlike the CT process. Our experiment shows that an efficient ET occurs from the 1Ls WSe$_2$-to-MoS$_2$ with an interlayer hBN, due to the resonant overlapping of the high-lying excitonic states between the two TMDs, resulting in enhanced HS MoS$_2$ PL emission. This type of unconventional ET from the *lower-to-higher* optical bandgap material is not typical in the TMD HSs. With increasing temperature, the ET process becomes weaker due to the increased electron-phonon scattering, destroying the enhanced MoS$_2$ emission. Our work provides new insight into the long-distance ET process and its effect on the photocarrier relaxation pathways.

**Introduction:**

Group-VI semiconducting transition metal dichalcogenides (TMDs) are formed by stacking of strongly bonded two-dimensional (2D) X-M-X layers (M = transition metals such as Mo, W etc. and X = chalcogens such as S, Se, and Te etc.), which are separated by the weakly bond interlayer van der Waals forces.[1] The first mechanical exfoliation of the monolayer (1L) molybdenum disulfide (MoS$_2$) film from a bulk crystal in 2010 led us to observe a strong photoluminescence (PL) emission[2,3] due to the indirect-to-direct bandgap transition from the bulk-to-1L regime.[4,5] Since then, researchers have been exploring exciting excitonic properties[6–11] in these 1L TMD materials. In particular, their strong light-matter interactions and high light absorption of up to ~15% in the solar spectrum[12] enabled researchers to realize the future prospects of 1L TMD-based optoelectronic device applications.[13] 2D heterostructures (HSs) made by the vertical stacking of different layered materials, have shown promising results for future ultrathin[14–16] and flexible[17] optoelectronic device applications. Recent advances in direct and patterned growth of 2D HSs[18,19] to obtain a clean large-area interface have also pushed the effort to make commercially available TMD-based device applications. However, one of the major challenges in commercializing the promised optoelectronic device applications is the lack of a comprehensive understanding in the competing interlayer processes, such as interlayer charge (CT) and energy transfer (ET) process, and their role in the photocarrier recombination mechanism.

CT and ET are the two main carrier relaxation pathways in the semiconductor HSs. The interlayer CT occurs due to an energy band offset in the HS[20] and the interlayer ET process happens when nonradiative energy from the excited donor material gets transferred to the acceptor material accompanied by a fluorescence emission from the acceptor material.[21,22] ET is observed as a reduction of the donor fluorescence intensity and carrier lifetime followed by an enhancement of the acceptor fluorescence intensity.[22] The interlayer CT can be stopped by placing a thin layer of dielectric material in between the two semiconductors. Britnell *et al.*[23] showed that only four atomic-layer thick hexagonal boron nitride (hBN) is sufficient as a dielectric medium to block the electron tunneling between the two graphene layers. Unlike the CT process, in TMD HSs the long-distance interlayer ET process can survive up to several tens of nm.[24,25] Separating the materials far apart from each other to stop the ET process is not practical for ultrathin device design. Thus, developing a comprehensive understanding of the long-distance interlayer ET process is an absolute necessity to create practical device applications.

In this work, we study the effect of resonant overlaps between the high-lying excitonic states of 1Ls tungsten diselenide ($WSe_2$) and $MoS_2$ on the interlayer ET process with a ~9 nm thick hBN charge-blocking layer. Both these TMD materials have overlapping higher energy B and C ($MoS_2$)/D ($WSe_2$) absorption features.[26,27] We show that resonant excitations at the $WSe_2$ B and D absorption regions results in $MoS_2$ PL enhancement in the HS area. We report that this PL enhancement is due to the interlayer ET process from the $WSe_2$-to-$MoS_2$ layer. This type of ET process from the *lower-to-higher* optical bandgap material is not typical in the TMD HSs, since ET conventionally happens from the higher-to-lower bandgap 2D materials.[28–33] In this work, we employ multiple optical spectroscopic techniques at cryogenic temperature (8 K); such as µ-PL, µ-photoluminescence excitation (PLE) and differential reflectance contrast (RC), complemented by the density functional theory (DFT) calculation of spin-resolved band structures to study the ET process. Our work reveals an unconventional interlayer ET process in the TMD HSs. This will significantly contribute to creating a comprehensive understanding of the long-range interlayer ET process and its role to influence the photocarrier radiative recombination processes in these semiconducting HSs.

**Results and discussion:**

Figure 1a shows the optical micrograph of the fabricated MoS$_2$-hBN-WSe$_2$ HS (see the methods for fabrication details). The inset of Figure 1a shows the schematic illustration of the cross-section of the sample. We introduce ~9 nm thick interlayer hBN (see Supplementary Figure S1) to eliminate any effect related to the interlayer CT in our system.[23] The optical absorption of the TMD materials reflects their single-electron energy band structure. The low temperature RC spectra (see methods for the details) measured at 8 K show strong overlaps between the B peaks of both materials and the WSe$_2$ D peak with the MoS$_2$ C peak (shaded areas in Figure 1b), which agrees well with the previously published reports.[26,27] In the later sections, we discuss how these strong overlaps help us to observe the reported ET from the *lower-to-higher* bandgap (WSe$_2$-to-MoS$_2$) material. The HS spectrum (Figure 1b) shows similar RC resonance positions as compared to the individual 1Ls, indicating no major strain-induced effect[34] in the HS area. A and B excitonic peaks occur due to the excitonic transitions at the K$^+$/K$^-$ valley in the *k*-space[2,3] and higher energy excitonic transitions, such as C and D, are the results of the 'band nesting'[35,36] in the Brillouin zone. 'Band-nested' regions occur due to the identical dispersion in the valence (VB) and conduction (CB) bands over a region in the Brillouin zone. For 1L MoS$_2$, both the VB maximum and the CB minimum are located at the K$^+$/K$^-$ point in the Brillouin zone. In the case of WSe$_2$, while the VB maximum is located at the K$^+$/K$^-$ point, the CB minimum is situated at the Λ point.[2,37] The 'band-nesting' region happens in between the Γ and Λ point.[35,36] Figure 1c shows the DFT calculated electronic band structures (see Supplementary Information for the details) along the Γ-K$^+$ direction in the Brillouin zone. For both the band structures, we match the optical bandgaps with the corresponding PL energies. All types of optical transitions are shown with different colors of arrows (Figure 1c).

PLE maps (see methods for the experimental details) taken at 8 K show the emission landscapes of the three individual areas (Figures 2a-2b, S2-S3). In Figure 2a, we saturate the WSe$_2$ emission intensity in order to visualize the MoS$_2$ emission. After comparing the MoS$_2$ emission intensities, we observe a significantly enhanced MoS$_2$ PL in the HS area as compared to the 1L region (Figures 2a-2b). The horizontal cuts at the excitation energies of 2.85 eV and 2.12 eV (black dotted lines in Figures 2a-2b) reveal that the $MoS_2$ PL emission in the HS is enhanced by a factor of ~1.9 and ~1.7, respectively as compared to the 1L area (Figures 2c-2d). The PL enhancement factor is defined here as the ratio of PL intensity in the HS area to the 1L area under the same excitation and accumulation conditions. Similarly, PLE (vertical cut along the 1.92 eV emission energy in Figures 2a-2b) shows an overall increase of the HS $MoS_2$ PL emission throughout the entire excitation range as compared to the 1L $MoS_2$ region (Figure 2e). It is important to mention that the total optical absorption in the HS area did not change much as compared to the each 1L areas (Figure 1b). However, the enhancement in the HS PLE (Figure 2e) suggests that the internal PL efficiency of the HS system was increased due to the ET process. We note that, the below-bandgap pronounced emission from the $MoS_2$ defect states (Figure 2b) is typical for the exfoliated and non-encapsulated samples.[38] We would also like to mention that in the HS PLE map we also see an enhancement in the $WSe_2$ excitonic emission (Figures S2-S3) due to the conventional ET from the $MoS_2$-to-$WSe_2$ layer. We did not include any discussion related to the enhancement of the $WSe_2$ PL in this study, as a similar type of ET has already been reported in a previous work.[28]

In this paragraph, we take into consideration of all other possible scenarios in the HS PL enhancement process. We rule out the possibility of the observed PL enhancement in the $MoS_2$ emission due to the interference of the backscattering light, because the entire measured $MoS_2$ area (including the HS) is placed onto the same hBN flake (inset of Figure 1a). 1L $WSe_2$ (thickness <1 nm) in the HS area cannot modulate the interference pattern considering the ~9 nm interlayer and thick substrate hBNs. We also rule out the possible contribution of ET from the hBN defect states[39] to the HS $MoS_2$ PL enhancement process, as the ET from the same hBN thickness cannot result in more HS PL emission as compared to the 1L $MoS_2$ region. In order to check the data reproducibility, we made another HS with different fabrication protocol and non-identical hBN thicknesses, and observed an enhanced $MoS_2$ PL emission in the HS area (see Supporting Information for the details and Figure S4). There is another possibility, that the emitted light from the $MoS_2$ layer could be reflected by the encapsulated $WSe_2$ at low-temperature,[40] increasing the PL enhancement only at the HS area. To verify this, we made a similar HS on the transparent ultraflat quartz substrate (Figure S5). For this transparent sample we observe shifts in the absorption peaks due to the change in the dielectric environment, which destroys the resonant overlap of the B-excitonic levels between the two materials. As a result, we see a quenching in the HS $MoS_2$ PL emission and an increase in the HS $WSe_2$ emission, proving that a one-way ET occurred from the $MoS_2$-to-$WSe_2$ layer.[28] This result proves that the reflection of the $MoS_2$ PL from the encapsulated $WSe_2$ layer has no effect here in the reported HS PL enhancement process. We conclude that the $MoS_2$ PL enhancement in the HS area is a result of an interlayer ET process from the $WSe_2$ layer.

Strong overlaps between the higher energy absorptions in both the investigated materials (Figure 1b) help us to study the effect of the interlayer ET process under those 'resonant' excitation conditions. The PL intensity map taken at 8 K under the excitation of 2.12 eV (B resonances overlap region) shows an overall enhanced $MoS_2$ emission in the HS area (Figure 3a). Similarly, an excitation at 2.85 eV energy ($WSe_2$ D and $MoS_2$ C peaks overlap region) shows an increased $MoS_2$ PL emission throughout the HS area (Figure 3b). Thus, proving that at both the excitation energies an efficient ET happened from the $WSe_2$-to-$MoS_2$ layer as discussed in the later section. The PL intensity maps (Figures 3a-3b) also show that the observed enhancement of the $MoS_2$ PL emission in the HS area is not a localized phenomenon. We note that although there is some non-uniformity in the HS PL intensity due to the typical inhomogeneous nature of the exfoliated samples, but the HS PL emission is always higher than the 1L $MoS_2$ area.

In order to study the effect of increasing temperature in our experiments, we performed PLE maps at 25 K, 100 K, and 200 K (Figure 4 and Figure S6). At 25 K, $MoS_2$ emissions in the HS area under both the excitation energies at ~2.83 eV and 2.2 eV show a similar enhancement factor of ~1.6 (Figure S7). These values are a slight reduction from the 8 K data. The PLE also shows a similar overall enhancement in the $MoS_2$ HS emission at 25 K (Figure 4c). Upon further increasing the temperature at 100 K and 200 K, we observe a complete vanishing of the $MoS_2$ PL enhancement in the HS (Figures 4d-4f). A slight quenching of the HS MoS$_2$ PLE at 100 K (Figure 4f) could be due to the conventional type-II HS ET[28] from the higher-to-lower bandgap material (MoS$_2$-to-WSe$_2$).

For MoS$_2$ and WSe$_2$, the schematics of the A and B transitions based on the VB and CB splitting are shown in Figure 5a. In these TMD monolayers, VB (VB1 and VB2) and CB (CB1 and CB2) spin splitting occurs due to the spin-orbit coupling and lack of inversion symmetry,[10,41] allowing possible absorptions based on the optical selection rule.[42,43] The corresponding PL emission, which comes from the direct radiative recombination at the optical bandgap, strongly depends on the spin-state of the CB (CB1 or CB2) electron and the VB (VB1 or VB2) hole at the K$^+$/K$^-$ point. Based on the allowed electron recombination from the CB1 or CB2 to the hole situated at the top of VB (VB2), the materials are divided into two categories; 'bright' or 'dark', respectively.[10] The calculated momentum-space energy landscape for the allowed optical transitions from VB2-to-CB1 and VB1-to-CB2 in the MoS$_2$ layer shows a smaller separation of ~150 meV at the K$^+$/K$^-$ point due to the spin splitting (Figures 5b-5c, Figure S8a), which matches well with the previous results.[44,45] WSe$_2$ shows a comparatively larger separation of ~500 meV at the K$^+$/K$^-$ point[46,47] for the VB2-to-CB2 and VB1-to-CB1 transitions (Figures 5d-5e, Figure S8b).

Optical excitation at the 'band-nested' region (MoS$_2$ C and WSe$_2$ D peak), excites electrons in the valley in between the Γ-Λ point in MoS$_2$ CB and around the Λ valley in WSe$_2$ CB. These excited photocarriers (electron and hole) instantly relax to their immediate band extreme points; Λ valley for electron and Γ hill for hole.[27] These carriers then further transfer to the band extrema *via* the extremely fast (<500 fs) intravalley scattering ($k_{iv}$).[48–50] In our HS, to describe the PL intensity map under the 2.12 eV excitation (Figure 3a), the only possible mechanism is shown as a schematic illustration in Figure 5f. Upon excitation with the 2.12 eV photons, photoexcited carries are generated at the WSe$_2$ B excitonic level. Due to the resonant overlap with the MoS$_2$ B level (Figure 1b), the WSe$_2$ B excitonic energy immediately transfer to the MoS$_2$ B and A bands, resulting in more carriers in the MoS$_2$ layer. The extra carriers at the MoS$_2$ B level transfer to the subsequent band extremum *via* an intervalley transition ($k_v$, i.e. B$_{K+/K-}$ to A$_{K-/K+}$ transitions), followed by a radiative recombination process ($k_r$) to the ground state (GS). Thus, we obtain an enhanced

MoS$_2$ PL emissions in the HS area with an excitation of 2.12 eV (Figure 3a). However, at an excitation energy of 2.85 eV (MoS$_2$ C and WSe$_2$ D peak overlap region, Figure 1b), two possible ET channels can play a crucial role. First, ET from the WSe$_2$ D level can directly generate more carriers at the MoS$_2$ C level due to the resonant overlapping. These extra carriers radiatively recombine at the band extremum *via* intravalley transition ($k_{iv}$), and giving rise to more MoS$_2$ PL emissions in the HS area, as shown in the schematic of Figure 5g (grey colored ET process). Another possibility is that upon the 2.85 eV excitation carriers generated at the WSe$_2$ D level scatter to the WSe$_2$ B level *via* the analogous $k_{iv}$ process and then transfer to the MoS$_2$ B and A level *via* ET process. Finally, it increases the MoS$_2$ PL emission similar to the 2.12 eV excitation process (black colored ET process in Figure 5g). Interestingly, an excitation at the WSe$_2$ C absorption peak (2.56 eV) does not result in any MoS$_2$ PL emission (Figure S9), indicating that interlayer coupling between the suitable levels was not possible at this excitation due to the immediate photoexcited carrier transfer to the WSe$_2$ A level. Hence, no enhancement in the MoS$_2$ HS PL emission due to the ET process is also apparent.

Our model to describe the enhanced MoS$_2$ PL emission from the HS area also supports the temperature-dependent data. Photocarriers go through a series of phonon scattering before relaxing to the ground state. At low temperature, electron-phonon scattering dominates.[51] With the increasing temperature, other types of scattering processes such as anharmonic phonon–phonon scattering and phonon structure scattering,[52] start to dominate. Thus, with the increasing temperature, the intravalley transition becomes weaker due to the multiple-phonon scattering and eventually a minor fraction of the photocarriers generated at the 'band-nested' region can be transferred to the K$^+$/K$^-$ point for radiative recombination. Furthermore, the thermal activation should make the 'hot' carrier transfer to the band extremum extremely faster (<100 fs),[53] preventing the coupling between the materials' corresponding energy levels. These eventually result in a complete disappearance of the MoS$_2$ PL enhancement in the HS area at higher temperatures (100 K and 200 K).

Considering the temperature-dependent data we can conclude that at higher excitation energy (~2.85 eV) the ET process *via* WSe$_2$ B to MoS$_2$ B and A level dominates (black colored ET process in Figure 5h) in our experiment. Otherwise, with increasing the temperature we should observe an enhanced MoS$_2$ HS PL emission. At cryogenic temperature, the fast intravalley scattering ($k_{iv}$) in TMDs occur at ~100-500 fs timescale.[48–50,53] Whereas, intervalley transitions ($k_v$) occur at a longer timescale of a few ps range.[54,55] Our study suggests that the reported ET happened at a faster timescale than the intervalley transition and slower than the intravalley transition. Otherwise, the ET from the lower optical bandgap WSe$_2$ cannot excite more carriers in the higher bandgap MoS$_2$, resulting in an enhanced HS MoS$_2$ PL emission. Finding the 'true' ET timescale in our experiment will require an ultrafast study, which is beyond the scope of this work. It is also important to mention that with the increasing temperature the effect of band renormalization in the ET process to alter the radiative recombination pathway of the photocarriers cannot be ignored. A thorough investigation of the band renormalization effect in the ET process is required in the future work.

In conclusion, our study shows that strong light matter interaction in the 1L MoS$_2$ and WSe$_2$ 'band-nested' region allows us to observe an unusual ET process from the *lower-to-higher* bandgap (WSe$_2$-to-MoS$_2$) material. The excitation-dependent PL intensity maps prove that the reported HS MoS$_2$ PL enhancement is not a localized phenomenon due to the materials local property change, the entire HS area shows this enhanced PL emission. Finally, the temperature-dependent study proves that with the increasing temperature due to the growing electron-phonon scattering, the carriers transfer to the band extremum become faster, preventing ET from the WSe$_2$ (smaller gap) to the MoS$_2$ (larger gap) layer. Our findings provide an important insight into the interlayer ET process in these layered materials and will help to build a comprehensive understanding about the competing interlayer processes for developing future TMD-based optoelectronic device applications.

**Methods:**

**HS fabrication**

We fabricated three HSs using two fabrication protocols. HSs in Figure 1a and S5, were fabricated using PDMS-based dry transfer technique at the University of Warsaw. Bottom hBN layer was directly cleaved on the SiO$_2$/Si substrate. MoS$_2$-hBN-WSe$_2$ layers were exfoliated onto the Gel-Pak (PDMS) films and were stacked layer-by-layer (in reverse order) onto each other using a home-built semiautomatic transfer stage. The other HS in Figure S4 was partially fabricated using a robotic fabrication system (QPress) at the Brookhaven National Laboratory (the details in Supporting Information). MoS$_2$, WSe$_2$ and hBN bulk crystals for exfoliation were obtained from the Graphene Supermarket, HQ Graphene and National Institute for Materials Science, respectively.

**Characterization**

We used Bruker Dimension Icon with NanoScope 6 controller in 'ScanAsyst' (peak force tapping) mode to obtain high resolution AFM image.

The differential RC measurements were performed using a super-continuum light source (without a monochromator) focused by a Nikon L Plan 100x (N.A. 0.7) objective and directed into a spectrometer. Sample was loaded in a cryostat and cooled with continuous flow of liquid helium (LHe). The differential reflectance is defined by $(R_s-R_{sub})/(R_s+R_{sub})$, where $R_s$ is the reflected light intensity from the TMD sample areas and $R_{sub}$ from the hBN/Si substrate.

We performed the µ-PL/PLE experiments by using a super-continuum light source coupled with a monochromator as an excitation source. The incident light was focused using a Mitutoyo M Plan 50x (N.A. 0.75) objective. The excitation power was constant throughout the measurements and the average power on the sample was kept ~50 µW (spot diameter ~1 µm) to avoid any high power induced nonlinear effects from the sample. For PLE experiment sample was loaded in a LHe cryostat to reach the minimum temperature of ~5 K during the experiments.

**Data availability:**

All the data necessary to conclude the results are presented in the manuscript and supplementary information.

**Code availability:**

The technical details of the theoretical calculations are available from the corresponding authors upon reasonable request.

**References:**

(1) Mattheiss, L. F. Band Structures of Transition-Metal-Dichalcogenide Layer Compounds. *Phys. Rev. B* **1973**, *8* (8), 3719–3740. https://doi.org/10.1103/PhysRevB.8.3719.

(2) Splendiani, A.; Sun, L.; Zhang, Y.; Li, T.; Kim, J.; Chim, C. Y.; Galli, G.; Wang, F. Emerging Photoluminescence in Monolayer MoS2. *Nano Lett.* **2010**, *10* (4), 1271–1275. https://doi.org/10.1021/nl903868w.

(3) Mak, K. F.; Lee, C.; Hone, J.; Shan, J.; Heinz, T. F. Atomically Thin MoS2: A New Direct-Gap Semiconductor. *Phys. Rev. Lett.* **2010**, *105* (13), 136805. https://doi.org/10.1103/PhysRevLett.105.136805.

(4) Jin, W.; Yeh, P.-C.; Zaki, N.; Zhang, D.; Sadowski, J. T.; Al-Mahboob, A.; van der Zande, A. M.; Chenet, D. A.; Dadap, J. I.; Herman, I. P.; Sutter, P.; Hone, J.; Osgood, R. M. Direct Measurement of the Thickness-Dependent Electronic Band Structure of MoS2 Using Angle-Resolved Photoemission Spectroscopy. *Phys. Rev. Lett.* **2013**, *111* (10), 106801. https://doi.org/10.1103/PhysRevLett.111.106801.

(5) Yeh, P. C.; Jin, W.; Zaki, N.; Zhang, D.; Liou, J. T.; Sadowski, J. T.; Al-Mahboob, A.; Dadap, J. I.; Herman, I. P.; Sutter, P.; Osgood, R. M. Layer-Dependent Electronic Structure of an Atomically Heavy Two-Dimensional Dichalcogenide. *Phys. Rev. B* **2015**, *91* (4), 041407. https://doi.org/10.1103/PhysRevB.91.041407.

(6) Xiao, D.; Liu, G. B.; Feng, W.; Xu, X.; Yao, W. Coupled Spin and Valley Physics in Monolayers of MoS2 and Other Group-VI Dichalcogenides. *Phys. Rev. Lett.* **2012**, *108* (19), 196802. https://doi.org/10.1103/PhysRevLett.108.196802.

(7) Mak, K. F.; He, K.; Lee, C.; Lee, G. H.; Hone, J.; Heinz, T. F.; Shan, J. Tightly Bound Trions in Monolayer MoS2. *Nature Materials* **2013**, *12* (3), 207–211. https://doi.org/10.1038/nmat3505.

(8) Ross, J. S.; Wu, S.; Yu, H.; Ghimire, N. J.; Jones, A. M.; Aivazian, G.; Yan, J.; Mandrus, D. G.; Xiao, D.; Yao, W.; Xu, X. Electrical Control of Neutral and Charged Excitons in a Monolayer Semiconductor. *Nature Communications* **2013**, *4* (1), 1474. https://doi.org/10.1038/ncomms2498.

(9) Mouri, S.; Miyauchi, Y.; Matsuda, K. Tunable Photoluminescence of Monolayer MoS2 via Chemical Doping. *Nano Lett.* **2013**, *13* (12), 5944–5948. https://doi.org/10.1021/nl403036h.

(10) Koperski, M.; Molas, M. R.; Arora, A.; Nogajewski, K.; Slobodeniuk, A. O.; Faugeras, C.; Potemski, M. Optical Properties of Atomically Thin Transition Metal Dichalcogenides: Observations and Puzzles. *Nanophotonics* **2017**, *6* (6), 1289–1308. https://doi.org/10.1515/nanoph-2016-0165.

(11) Man, M. K. L.; Madéo, J.; Sahoo, C.; Xie, K.; Campbell, M.; Pareek, V.; Karmakar, A.; Wong, E. L.; Al-Mahboob, A.; Chan, N. S.; Bacon, D. R.; Zhu, X.; Abdelrasoul, M. M. M.; Li, X.; Heinz, T. F.; da Jornada, F. H.; Cao, T.; Dani, K. M. Experimental Measurement of the Intrinsic Excitonic Wave Function. *Science Advances* *7* (17), eabg0192. https://doi.org/10.1126/sciadv.abg0192.

(12) Wurstbauer, U.; Miller, B.; Parzinger, E.; Holleitner, A. W. Light–Matter Interaction in Transition Metal Dichalcogenides and Their Heterostructures. *Journal of Physics D: Applied Physics* **2017**, *50* (17), 173001. https://doi.org/10.1088/1361-6463/aa5f81.

(13) Eda, G.; Maier, S. A. Two-Dimensional Crystals: Managing Light for Optoelectronics. *ACS Nano* **2013**, *7* (7), 5660–5665. https://doi.org/10.1021/nn403159y.

(14) Britnell, L.; Ribeiro, R. M.; Eckmann, A.; Jalil, R.; Belle, B. D.; Mishchenko, A.; Kim, Y. J.; Gorbachev, R. V.; Georgiou, T.; Morozov, S. V.; Grigorenko, A. N.; Geim, A. K.; Casiraghi, C.;

Neto, A. H. C.; Novoselov, K. S. Strong Light-Matter Interactions in Heterostructures of Atomically Thin Films. *Science* **2013**, *340* (6138), 1311–1314. https://doi.org/10.1126/science.1235547.

(15) Huo, N.; Tongay, S.; Guo, W.; Li, R.; Fan, C.; Lu, F.; Yang, J.; Li, B.; Li, Y.; Wei, Z. Novel Optical and Electrical Transport Properties in Atomically Thin WSe2/MoS2 p–n Heterostructures. *Advanced Electronic Materials* **2015**, *1* (5), 1400066. https://doi.org/10.1002/aelm.201400066.

(16) Huo, N.; Kang, J.; Wei, Z.; Li, S. S.; Li, J.; Wei, S. H. Novel and Enhanced Optoelectronic Performances of Multilayer MoS2–WS2 Heterostructure Transistors. *Advanced Functional Materials* **2014**, *24* (44), 7025–7031. https://doi.org/10.1002/adfm.201401504.

(17) Gao, L. Flexible Device Applications of 2D Semiconductors. *Small* **2017**, *13* (35), 1603994. https://doi.org/10.1002/smll.201603994.

(18) Liang, J.; Zhu, X.; Chen, M.; Duan, X.; Li, D.; Pan, A. Controlled Growth of Two-Dimensional Heterostructures: In-Plane Epitaxy or Vertical Stack. *Acc. Mater. Res.* **2022**, *3* (10), 999–1010. https://doi.org/10.1021/accountsmr.2c00096.

(19) Lu, F.; Karmakar, A.; Shahi, S.; Einarsson, E. Selective and Confined Growth of Transition Metal Dichalcogenides on Transferred Graphene. *RSC Adv.* **2017**, *7* (59), 37310–37314. https://doi.org/10.1039/C7RA07772F.

(20) Bradac, C., Xu, Z. Q. & Aharonovich, I. Quantum Energy and Charge Transfer at Two-Dimensional Interfaces. *Nano Lett.* **21**, 1193–1204 (2021).

(21) Förster, T. Energy Migration and Fluorescence. *Journal of Biomedical Optics* **2012**, *17* (1), 011002. https://doi.org/10.1117/1.JBO.17.1.011002.

(22) Energy Transfer. In *Principles of Fluorescence Spectroscopy*; Lakowicz, J. R., Ed.; Springer US: Boston, MA, 2006; pp 443–475. https://doi.org/10.1007/978-0-387-46312-4_13.

(23) Britnell, L.; Gorbachev, R. V.; Jalil, R.; Belle, B. D.; Schedin, F.; Katsnelson, M. I.; Eaves, L.; Morozov, S. V.; Mayorov, A. S.; Peres, N. M. R.; Castro Neto, A. H.; Leist, J.; Geim, A. K.; Ponomarenko, L. A.; Novoselov, K. S. Electron Tunneling through Ultrathin Boron Nitride Crystalline Barriers. *Nano Lett.* **2012**, *12* (3), 1707–1710. https://doi.org/10.1021/nl3002205.

(24) Federspiel, F.; Froehlicher, G.; Nasilowski, M.; Pedetti, S.; Mahmood, A.; Doudin, B.; Park, S.; Lee, J. O.; Halley, D.; Dubertret, B.; Gilliot, P.; Berciaud, S. Distance Dependence of the Energy Transfer Rate from a Single Semiconductor Nanostructure to Graphene. *Nano Lett.* **2015**, *15* (2), 1252–1258. https://doi.org/10.1021/nl5044192.

(25) Karmakar, A.; Al-Mahboob, A.; Petoukhoff, C. E.; Kravchyna, O.; Chan, N. S.; Taniguchi, T.; Watanabe, K.; Dani, K. M. Dominating Interlayer Resonant Energy Transfer in Type-II 2D Heterostructure. *ACS Nano* **2022**, *16* (3), 3861–3869. https://doi.org/10.1021/acsnano.1c08798.

(26) Li, Y.; Chernikov, A.; Zhang, X.; Rigosi, A.; Hill, H. M.; van der Zande, A. M.; Chenet, D. A.; Shih, E. M.; Hone, J.; Heinz, T. F. Measurement of the Optical Dielectric Function of Monolayer Transition-Metal Dichalcogenides: MoS2, MoSe2, WS2, and WSe2. *Phys. Rev. B* **2014**, *90* (20), 205422. https://doi.org/10.1103/PhysRevB.90.205422.

(27) Kozawa, D.; Kumar, R.; Carvalho, A.; Kumar Amara, K.; Zhao, W.; Wang, S.; Toh, M.; Ribeiro, R. M.; Castro Neto, A. H.; Matsuda, K.; Eda, G. Photocarrier Relaxation Pathway in Two-Dimensional Semiconducting Transition Metal Dichalcogenides. *Nature Communications* **2014**, *5* (1), 4543. https://doi.org/10.1038/ncomms5543.

(28) Kozawa, D.; Carvalho, A.; Verzhbitskiy, I.; Giustiniano, F.; Miyauchi, Y.; Mouri, S.; Castro Neto, A. H.; Matsuda, K.; Eda, G. Evidence for Fast Interlayer Energy Transfer in MoSe2/WS2 Heterostructures. *Nano Lett.* **2016**, *16* (7), 4087–4093. https://doi.org/10.1021/acs.nanolett.6b00801.

(29) Zhang, Q.; Linardy, E.; Wang, X.; Eda, G. Excitonic Energy Transfer in Heterostructures of Quasi-2D Perovskite and Monolayer WS2. *ACS Nano* **2020**, *14* (9), 11482–11489. https://doi.org/10.1021/acsnano.0c03893.

(30) Froehlicher, G.; Lorchat, E.; Berciaud, S. Charge Versus Energy Transfer in Atomically Thin Graphene-Transition Metal Dichalcogenide van Der Waals Heterostructures. *Phys. Rev. X* **2018**, *8* (1), 011007. https://doi.org/10.1103/PhysRevX.8.011007.

(31) Ferrante, C.; Di Battista, G.; López, L. E. P.; Batignani, G.; Lorchat, E.; Virga, A.; Berciaud, S.; Scopigno, T. Picosecond Energy Transfer in a Transition Metal Dichalcogenide–Graphene

Heterostructure Revealed by Transient Raman Spectroscopy. *Proceedings of the National Academy of Sciences* **2022**, *119* (15), e2119726119. https://doi.org/10.1073/pnas.2119726119.

(32) Wu, L.; Chen, Y.; Zhou, H.; Zhu, H. Ultrafast Energy Transfer of Both Bright and Dark Excitons in 2D van Der Waals Heterostructures Beyond Dipolar Coupling. *ACS Nano* **2019**, *13* (2), 2341–2348. https://doi.org/10.1021/acsnano.8b09059.

(33) Hu, Z.; Hernández-Martínez, P. L.; Liu, X.; Amara, M. R.; Zhao, W.; Watanabe, K.; Taniguchi, T.; Demir, H. V.; Xiong, Q. Trion-Mediated Förster Resonance Energy Transfer and Optical Gating Effect in WS2/HBN/MoSe2 Heterojunction. *ACS Nano* **2020**, *14* (10), 13470–13477. https://doi.org/10.1021/acsnano.0c05447.

(34) Mennel, L.; Smejkal, V.; Linhart, L.; Burgdörfer, J.; Libisch, F.; Mueller, T. Band Nesting in Two-Dimensional Crystals: An Exceptionally Sensitive Probe of Strain. *Nano Lett.* **2020**, *20* (6), 4242–4248. https://doi.org/10.1021/acs.nanolett.0c00694.

(35) Carvalho, A.; Ribeiro, R. M.; Castro Neto, A. H. Band Nesting and the Optical Response of Two-Dimensional Semiconducting Transition Metal Dichalcogenides. *Phys. Rev. B* **2013**, *88* (11), 115205. https://doi.org/10.1103/PhysRevB.88.115205.

(36) Liu, H.; Lu, J. Exciton Dynamics in Tungsten Dichalcogenide Monolayers. *Phys. Chem. Chem. Phys.* **2017**, *19* (27), 17877–17882. https://doi.org/10.1039/C7CP02510F.

(37) Jiang, H. Electronic Band Structures of Molybdenum and Tungsten Dichalcogenides by the GW Approach. *J. Phys. Chem. C* **2012**, *116* (14), 7664–7671. https://doi.org/10.1021/jp300079d.

(38) Molas, M. R.; Gołasa, K.; Bala, Ł.; Nogajewski, K.; Bartos, M.; Potemski, M.; Babiński, A. Tuning Carrier Concentration in a Superacid Treated MoS2 Monolayer. *Scientific Reports* **2019**, *9* (1), 1989. https://doi.org/10.1038/s41598-018-38413-6.

(39) Li, C.; Xu, Z. Q.; Mendelson, N.; Kianinia, M.; Wan, Y.; Toth, M.; Aharonovich, I.; Bradac, C. Resonant Energy Transfer between Hexagonal Boron Nitride Quantum Emitters and Atomically Layered Transition Metal Dichalcogenides. *2D Materials* **2020**, *7* (4), 045015. https://doi.org/10.1088/2053-1583/aba332.

(40) Scuri, G.; Zhou, Y.; High, A. A.; Wild, D. S.; Shu, C.; De Greve, K.; Jauregui, L. A.; Taniguchi, T.; Watanabe, K.; Kim, P.; Lukin, M. D.; Park, H. Large Excitonic Reflectivity of Monolayer MoSe2 Encapsulated in Hexagonal Boron Nitride. *Phys. Rev. Lett.* **2018**, *120* (3), 037402. https://doi.org/10.1103/PhysRevLett.120.037402.

(41) Kośmider, K.; González, J. W.; Fernández-Rossier, J. Large Spin Splitting in the Conduction Band of Transition Metal Dichalcogenide Monolayers. *Phys. Rev. B* **2013**, *88* (24), 245436. https://doi.org/10.1103/PhysRevB.88.245436.

(42) Molas, M. R.; Slobodeniuk, A. O.; Kazimierczuk, T.; Nogajewski, K.; Bartos, M.; Kapuściński, P.; Oreszczuk, K.; Watanabe, K.; Taniguchi, T.; Faugeras, C.; Kossacki, P.; Basko, D. M.; Potemski, M. Probing and Manipulating Valley Coherence of Dark Excitons in Monolayer WSe2. *Phys. Rev. Lett.* **2019**, *123* (9), 096803. https://doi.org/10.1103/PhysRevLett.123.096803.

(43) Robert, C.; Han, B.; Kapuscinski, P.; Delhomme, A.; Faugeras, C.; Amand, T.; Molas, M. R.; Bartos, M.; Watanabe, K.; Taniguchi, T.; Urbaszek, B.; Potemski, M.; Marie, X. Measurement of the Spin-Forbidden Dark Excitons in MoS2 and MoSe2 Monolayers. *Nature Communications* **2020**, *11* (1), 4037. https://doi.org/10.1038/s41467-020-17608-4.

(44) Eginligil, M.; Cao, B.; Wang, Z.; Shen, X.; Cong, C.; Shang, J.; Soci, C.; Yu, T. Dichroic Spin–Valley Photocurrent in Monolayer Molybdenum Disulphide. *Nature Communications* **2015**, *6* (1), 7636. https://doi.org/10.1038/ncomms8636.

(45) Kadantsev, E. S.; Hawrylak, P. Electronic Structure of a Single MoS2 Monolayer. *Solid State Communications* **2012**, *152* (10), 909–913. https://doi.org/10.1016/j.ssc.2012.02.005.

(46) Zhang, Y.; Ugeda, M. M.; Jin, C.; Shi, S. F.; Bradley, A. J.; Martín-Recio, A.; Ryu, H.; Kim, J.; Tang, S.; Kim, Y.; Zhou, B.; Hwang, C.; Chen, Y.; Wang, F.; Crommie, M. F.; Hussain, Z.; Shen, Z. X.; Mo, S. K. Electronic Structure, Surface Doping, and Optical Response in Epitaxial WSe2 Thin Films. *Nano Lett.* **2016**, *16* (4), 2485–2491. https://doi.org/10.1021/acs.nanolett.6b00059.

(47) Madéo, J.; Man, M. K. L.; Sahoo, C.; Campbell, M.; Pareek, V.; Wong, E. L.; Al-Mahboob, A.; Chan, N. S.; Karmakar, A.; Mariserla, B. M. K.; Li, X.; Heinz, T. F.; Cao, T.; Dani, K. M. Directly

Visualizing the Momentum-Forbidden Dark Excitons and Their Dynamics in Atomically Thin Semiconductors. *Science* **2020**, *370* (6521), 1199–1204. https://doi.org/10.1126/science.aba1029.

(48) Liu, F.; Li, Q.; Zhu, X. Y. Direct Determination of Momentum-Resolved Electron Transfer in the Photoexcited van Der Waals Heterobilayer WS2/MoS2. *Phys. Rev. B* **2020**, *101* (20), 201405. https://doi.org/10.1103/PhysRevB.101.201405.

(49) Shi, H.; Yan, R.; Bertolazzi, S.; Brivio, J.; Gao, B.; Kis, A.; Jena, D.; Xing, H. G.; Huang, L. Exciton Dynamics in Suspended Monolayer and Few-Layer MoS2 2D Crystals. *ACS Nano* **2013**, *7* (2), 1072–1080. https://doi.org/10.1021/nn303973r.

(50) Sim, S.; Park, J.; Song, J. G.; In, C.; Lee, Y. S.; Kim, H.; Choi, H. Exciton Dynamics in Atomically Thin MoS2: Interexcitonic Interaction and Broadening Kinetics. *Phys. Rev. B* **2013**, *88* (7), 075434. https://doi.org/10.1103/PhysRevB.88.075434.

(51) Liao, B.; Qiu, B.; Zhou, J.; Huberman, S.; Esfarjani, K.; Chen, G. Significant Reduction of Lattice Thermal Conductivity by the Electron-Phonon Interaction in Silicon with High Carrier Concentrations: A First-Principles Study. *Phys. Rev. Lett.* **2015**, *114* (11), 115901. https://doi.org/10.1103/PhysRevLett.114.115901.

(52) He, G. C.; Shi, L. N.; Hua, Y. L.; Zhu, X. L. The Phonon Scattering Mechanism and Its Effect on the Temperature Dependent Thermal and Thermoelectric Properties of a Silver Nanowire. *Phys. Chem. Chem. Phys.* **2022**, *24* (5), 3059–3065. https://doi.org/10.1039/D1CP04914C.

(53) Selig, M.; Berghäuser, G.; Raja, A.; Nagler, P.; Schüller, C.; Heinz, T. F.; Korn, T.; Chernikov, A.; Malic, E.; Knorr, A. Excitonic Linewidth and Coherence Lifetime in Monolayer Transition Metal Dichalcogenides. *Nature Communications* **2016**, *7* (1), 13279. https://doi.org/10.1038/ncomms13279.

(54) Song, Y.; Dery, H. Transport Theory of Monolayer Transition-Metal Dichalcogenides through Symmetry. *Phys. Rev. Lett.* **2013**, *111* (2), 026601. https://doi.org/10.1103/PhysRevLett.111.026601.

(55) Glazov, M. M.; Amand, T.; Marie, X.; Lagarde, D.; Bouet, L.; Urbaszek, B. Exciton Fine Structure and Spin Decoherence in Monolayers of Transition Metal Dichalcogenides. *Phys. Rev. B* **2014**, *89* (20), 201302. https://doi.org/10.1103/PhysRevB.89.201302.

**Acknowledgements:**

The work has been supported by the National Science Centre, Poland (Grant No. 2017/27/B/ST3/00205 and 2018/31/B/ST3/02111). K.W. and T.T. acknowledge support from the JSPS KAKENHI (Grant No. 19H05790, 20H00354 and 21H05233). This research used Quantum material press (QPress) of the Center for Functional Nanomaterials (CFN), which is a U.S. Department of Energy Office of Science User Facility, at Brookhaven National Laboratory under Contract No. DE-SC0012704. Authors acknowledge the help received from the research staffs at the Center of New Technologies (CeNT) in University of Warsaw.

**Author contributions:**

A.K. and A.A.M. conceived the project. A.K., A.A.M. and M.R.M. designed the experiments. A.K., S.P. and H.J. fabricated the samples. T.K., A.K., I.A., M.R. and M.R.M. performed the experiments. A.K. and A.A.M. analyzed the data. A.A.M. performed the theoretical calculations. A.K., A.A.M., M.R.M. and A.B. interpreted the results. T.T. and K.W. provided the bulk hBN for exfoliation. A.K. wrote the manuscript with feedback taken from all the coauthors.

**Competing interests:**

Authors declare no competing financial interests.

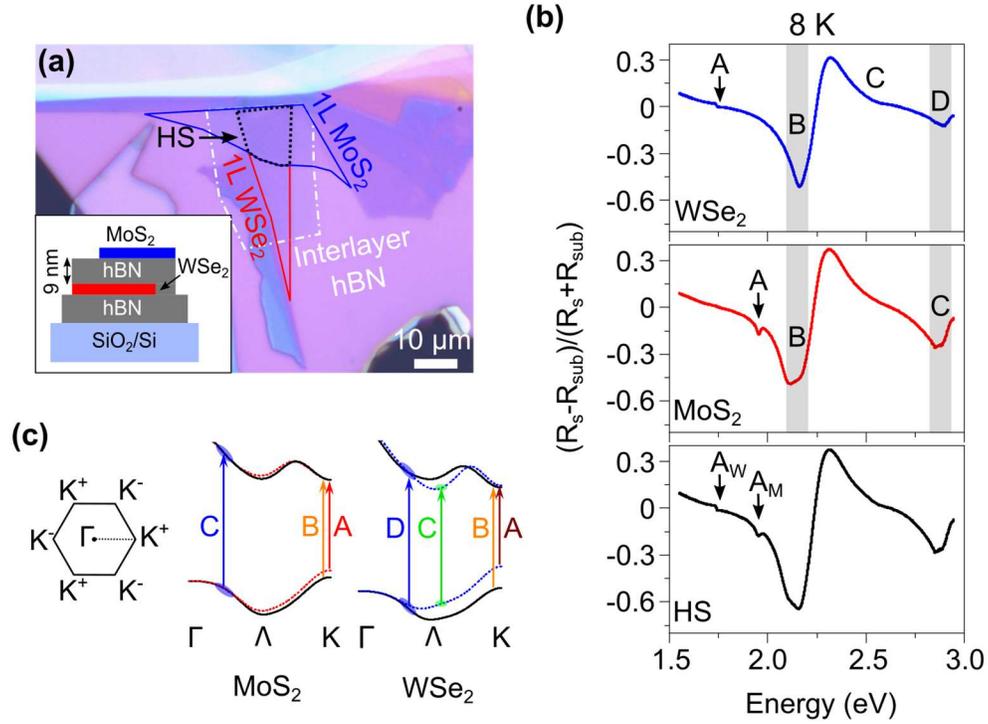

*Figure 1: (a)* Optical micrograph of the HS. Inset is the schematic illustration of the sample cross-section. The entire $MoS_2$ layer is placed on the same hBN thickness. *(b)* Differential reflectance contrast (RC) spectra from the three areas on the sample taken at 8 K. Shaded areas indicate the higher energy excitonic resonances between $MoS_2$ and $WSe_2$. HS shows the characteristics lower energy absorptions from both the $WSe_2$ ($A_W$) and $MoS_2$ ($A_M$) layer. *(c)* Single particle band structure of $MoS_2$ and $WSe_2$ along the $\Gamma$-K direction indicating the different optical transitions. Optical bandgaps were matched with the PL energies. C and D absorption peaks are the results of the 'band-nesting' in the Brillouin zone.

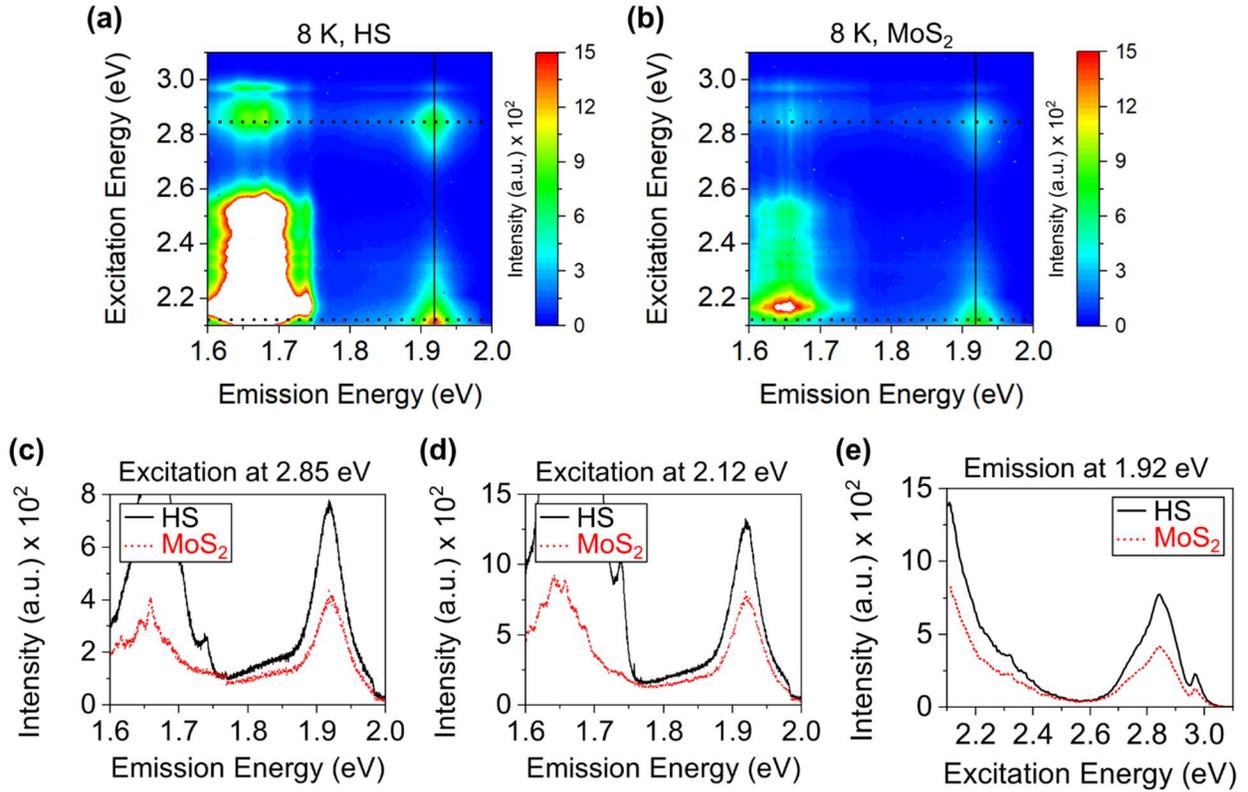

***Figure 2: (a)-(b)*** *PLE maps of the HS and MoS$_2$ area with the same intensity range taken at 8 K. WSe$_2$ emission intensity in the HS map is kept saturated to visualize the MoS$_2$ emission. MoS$_2$ shows a pronounced emission in the HS area. **(c)-(d)** (MoS$_2$ in) HS and MoS$_2$ PL emission intensities at 2.85 eV and 2.12 eV excitation energies, respectively (along the horizontal dotted lines in Figures 2(a)-(b)). Under both the excited energies, MoS$_2$ emissions in the HS are significantly enhanced as compared to the 1L area. **(e)** Comparison of HS and MoS$_2$ excitation profile at 1.92 eV emission energy (along the vertical solid lines in Figures 2(a)-(b)). Overall MoS$_2$ shows enhanced PLE intensity in the HS area.*

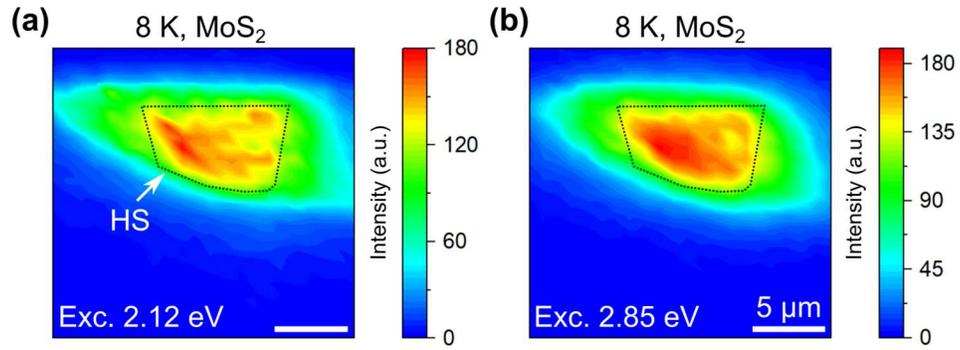

*Figure 3: (a)-(b)* MoS$_2$ photoluminescence (PL) intensity maps at 8 K under 2.12 eV and 2.85 eV excitation energy, respectively. MoS$_2$ emission in the HS area shows an overall increased PL emission as compared to the 1L region. The scale bars represent 5 µm length.

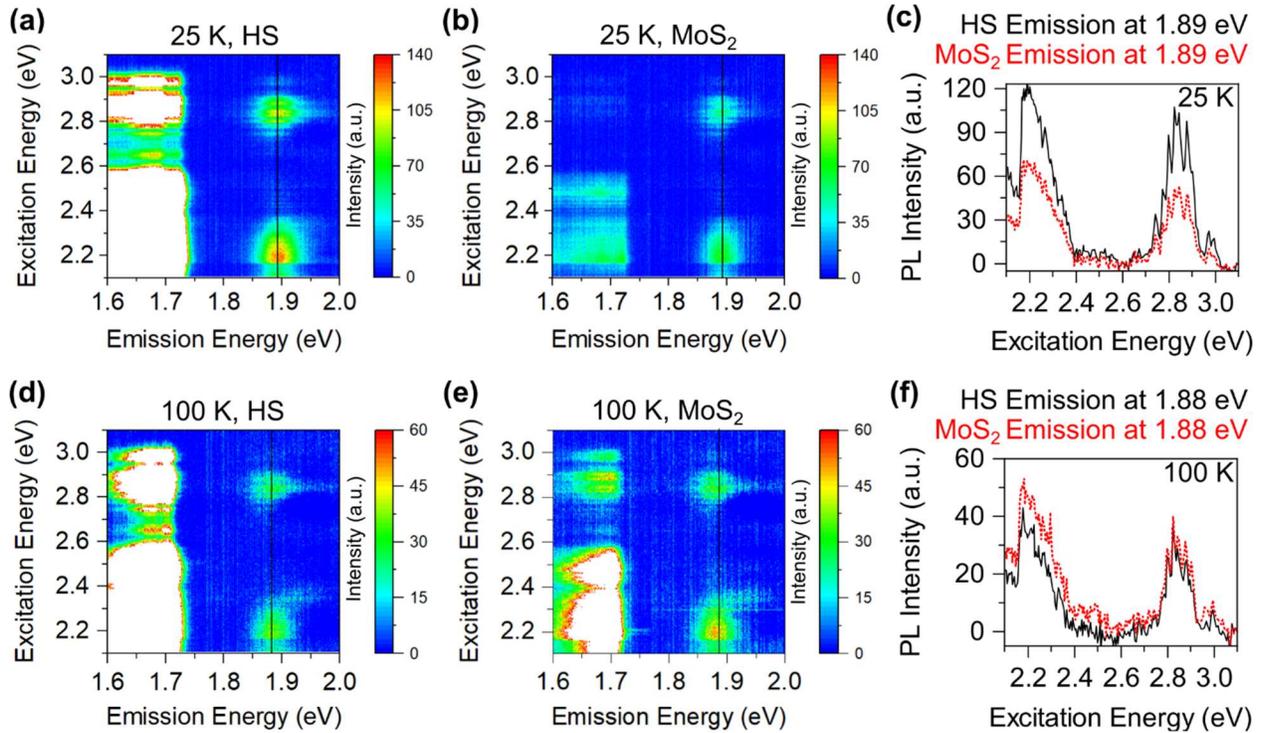

*Figure 4: (a)-(b)* HS and MoS$_2$ PLE maps at 25 K. MoS$_2$ PL emission in the HS area shows an enhancement as compared to the 1L area. *(c)* HS and MoS$_2$ PLE comparison along the vertical lines in (a)-(b). HS shows a slightly reduced MoS$_2$ PLE enhancement as compared to the 8 K map. *(d)-(e)* HS and MoS$_2$ PLE maps taken at 100 K. *(f)* Similar HS and MoS$_2$ PLE comparison at 100 K. MoS$_2$ in the HS area does not show any intensity enhancement at 100 K as compared to the 1L area. In all the HS maps, WSe$_2$ emission intensities are kept saturated to visualize the MoS$_2$ emission.
22

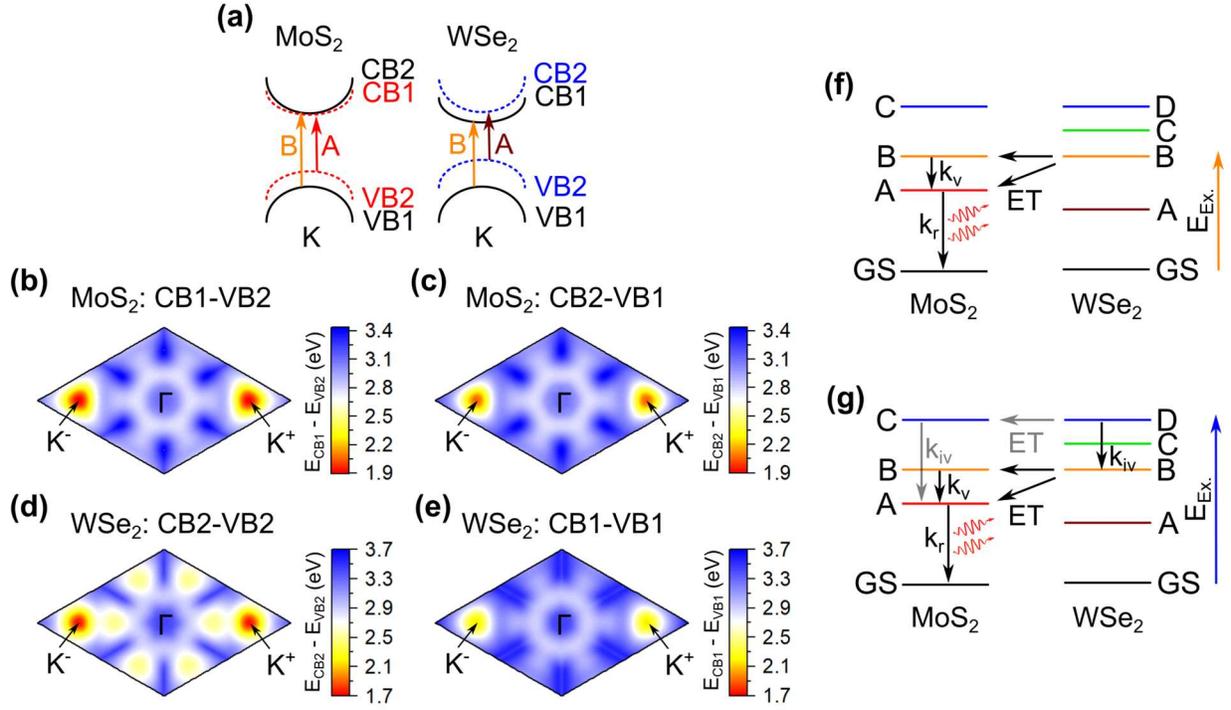

*Figure 5: (a)* Schematic illustration of the valence (VB) and conduction band (CB) splitting at the K valley in $MoS_2$ and $WSe_2$, respectively. *(b)-(c)* Calculated $MoS_2$ optical transitions along the $K^-$-$\Gamma$-$K^+$ direction from VB2 to CB1 and VB1 to CB2 (as shown in (a)), respectively. *(d)-(e)* Similar calculated $WSe_2$ momentum-space energy landscape along the $K^-$-$\Gamma$-$K^+$ direction from VB2 to CB2 and VB1 to CB1 (as shown in (a)), respectively. *(f)-(g)* Schematic illustration of the photocarrier relaxation pathways from the higher energy levels to the ground state (GS) in $MoS_2$ due to the energy transfer (ET) from $WSe_2$ after resonant excitation at ($WSe_2$) B and D excitonic level, respectively. Different types of transition are shown in the $MoS_2$ layer; such as intravalley scattering ($k_{iv}$), intervalley transition ($k_v$), and radiative recombination ($k_r$).

# SUPPORTING INFORMATION

# Excitation-Dependent High-Lying Excitonic Exchange *via* Interlayer Energy Transfer from *Lower-to-Higher* Bandgap 2D Material

*Arka Karmakar[1]\*, Tomasz Kazimierczuk[1], Igor Antoniazzi[1], Mateusz Raczyński[1], Suji Park[2], Houk Jang[2], Takashi Taniguchi[3], Kenji Watanabe[4], Adam Babiński[1], Abdullah Al-Mahboob[2ǂ], Maciej R. Molas[1#]*

[1] Division of Solid State Physics, Institute of Experimental Physics, Faculty of Physics, University of Warsaw, Pasteura 5, 02-093 Warsaw, Poland

[2] Center for Functional Nanomaterials, Brookhaven National Laboratory, Upton, NY 11973, USA

[3] International Center for Materials Nanoarchitectonics, National Institute for Materials Science, 1-1 Namiki, Tsukuba, Ibaraki 305-0044, Japan

[4] Research Center for Functional Materials, National Institute for Materials Science, 1-1 Namiki, Tsukuba, Ibaraki 305-0044, Japan

\* arka.karmakar@fuw.edu.pl

ǂ aalmahboo@bnl.gov

# maciej.molas@fuw.edu.pl

**Details of the theoretical calculations:**

We computed the ground state band structure of 1Ls $MoS_2$ and $WSe_2$ employing the density functional theory (DFT) calculations using the Materials Studio CASTEP (CAmbridge Serial Total Energy Package) version 2021 HF1, *ab initio* Total Energy Program (first principles methods using CASTEP).[1] Prior to the band structure calculation, we performed the geometry optimization (GO) for the bulk crystal structure using DFT-D (GGA + dispersion correction) method - Perdew-Bruke-Ernzerhof (PBE) GGA functional[2] along with the dispersion correction (van der Waals correction accounted employing the dispersion correction for DFT) by Tkatchenko-Scheffler (TS) method,[3] which was performed using the DFT Semi-Empirical Dispersion Interaction Correction (DFT-SEDC) module.[4] We obtained the electron relativistic correction using the DSPP (DFT-Semicore Pseudopotential).[5] During the GO of the bulk structure, symmetry constrained was imposed considering the International Table #194 (hexagonal, symmetry group P6$_3$/mmc, crystal class 6/mmm) for the bulk $MoS_2$ and $WSe_2$. Following the bulk geometry optimization, crystal was cleaved parallel to the layer (c* terminated) and then a vacuum slab > 20 Å was added along the c* to make the 1L TMD structures. Final GO for the atomic arrangement within the 1L and the in-plane lattice parameters were further optimized constraining the 2D lattice symmetry employing the identical GGA functional and dispersion correction as above but also including the spin-orbit coupling in the total energy calculations. In order to include the spin-orbit coupling, norm-conserving potentials in CASTEP were generated using the kinetic energy optimization scheme developed by Lin *et al.*[6] The spin orbit coupling was included using the j-dependent pseudopotentials developed for CASTEP based on the work by *ref*.[7] Following the final step of GO, band structure calculation was performed considering the ultra-fine k-spacing (k-spacing in single point energy calculation corresponding to 50x50x1 supercell or better and spectral k-spacing of 0.0005Å-1).

After computation of the electronic band structure in CASTEP, scissors have applied to the band structure plot to match with the bandgap obtained from the PL spectroscopy measurements.

**AFM data of the interlayer hBN thickness of the main MoS$_2$-hBN-WSe$_2$ sample (Figure S1):**

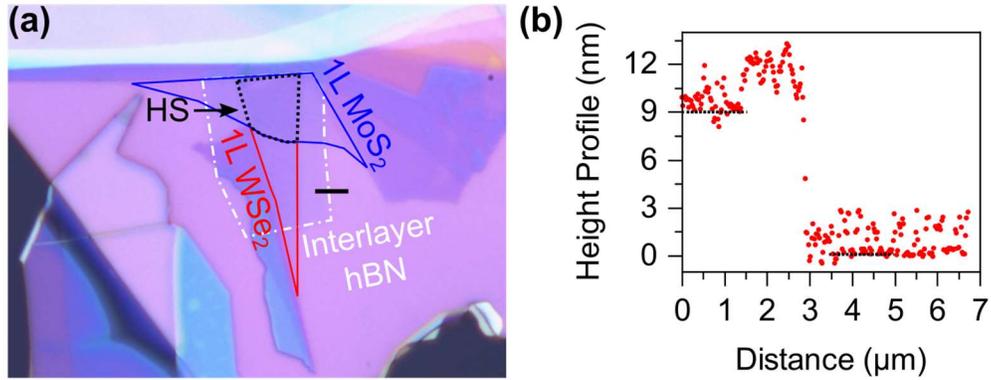

*Figure S1: (a) Optical micrograph of the HS. Solid black line indicates the line region of the AFM height profile. (b) AFM height profile of the interlayer hBN shows the thickness of ~9 nm.*

**Full PLE map of the HS and 1L WSe$_2$ region from the main MoS$_2$-hBN-WSe$_2$ sample (Figure S2-S3):**

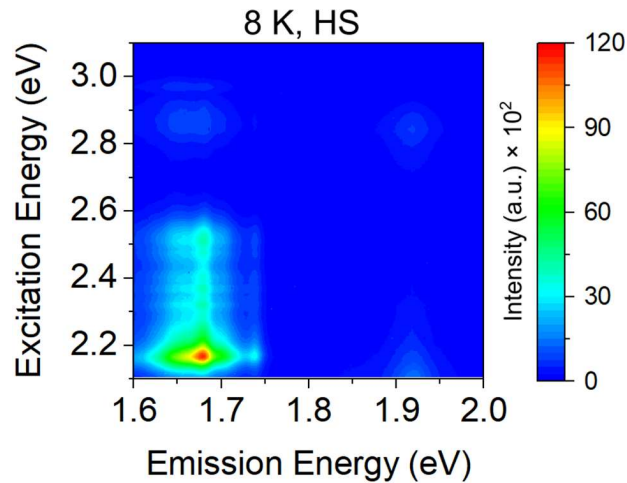

*Figure S2: HS PLE map taken at 8K to visualize the WSe$_2$ emission.*

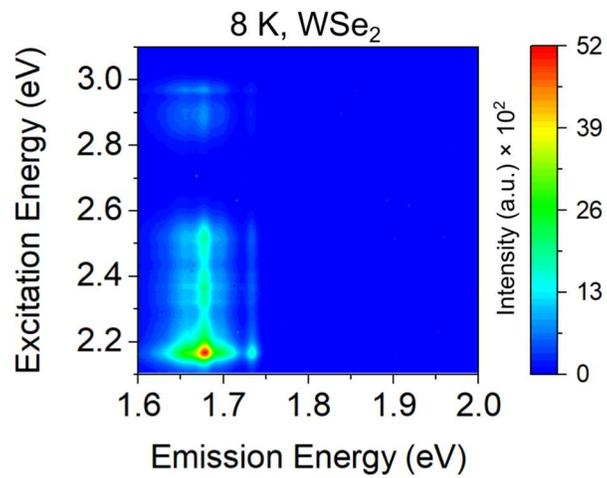

*Figure S3: WSe₂ PLE map taken at 8K. WSe2 excitonic intensity is reduced as compared to the HS area (Figure S2).*

**HS fabrication using Quantum material press (QPress) facility at Brookhaven National Laboratory (Figure S4):**

Each layer (except $MoS_2$) was exfoliated on $SiO_2$/Si substrates by a roll-to-roll exfoliation machine (or R2R exfoliator) in QPress. Exfoliation conditions were controlled following the recipes developed and provided by the QPress facility. We transferred the exfoliated flaked by a dry transfer method using a stamp consisting of a thin polycarbonate (PC) film onto a polydimethylsiloxane (PDMS) dome mounted on a glass slide. Unlike PDMS-based transfer method (bottom-up), the interlayer hBN was picked up by the PC/PDMS stamp first and then $WSe_2$ and bottom hBN was picked up sequentially (top-down). When picking-up, the substrate was heated up to ~150 °C. The (interlayer) hBN-$WSe_2$-(bottom) hBN structure stacked under the PC/PDMS stamp was released onto the final $SiO_2$/Si substrate by melting the PC layer at ~180 °C. PC residues were rinsed by chloroform, acetone and isopropanol. We soft baked the sample before transferring the top $MoS_2$ layer using the PDMS-based dry transfer technique as mentioned in the main text. We did not complete the entire HS using the PC/PDMS stamp to protect the top $MoS_2$ layer from chemical doping during the chloroform, acetone and isopropanol treatment.

Due to an instrumental limitation the PLE measurements were performed in the range of ~2.12-2.5 eV excitation energies with an average power ~ 40 µW. Figure S4 shows the optical image and PLE maps of the sample.

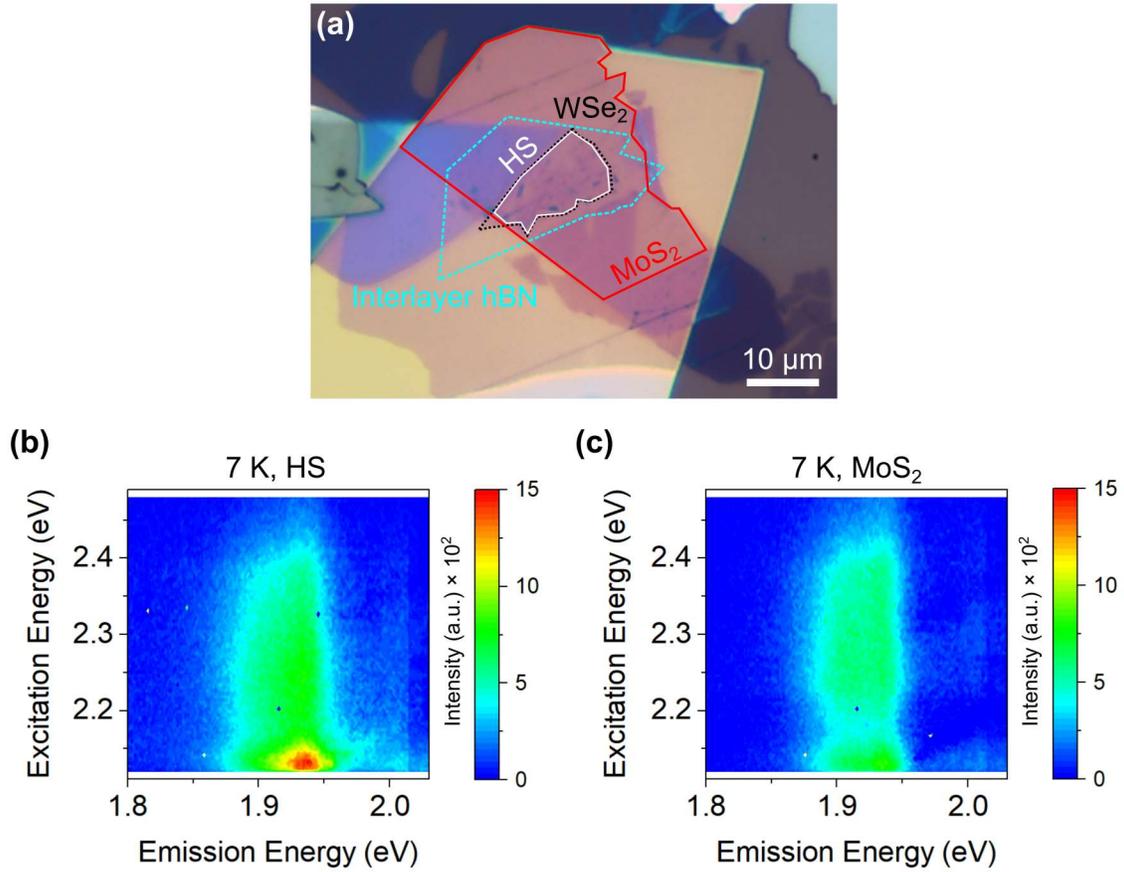

*Figure S4: (a) Optical micrograph of the $MoS_2$-hBN-$WSe_2$ HS. (b)-(c) PLE maps of the HS and $MoS_2$ taken at 7 K, respectively. $MoS_2$ PL emission shows an increased intensity in the HS area. Both the plots have the same intensity range.*

**MoS$_2$-hBN-WSe$_2$ HS on transparent quartz substrate (Figure S5):**

This HS on the ultraflat transparent quartz substrate was fabricated using the PDMS-based transfer technique as described in the method section of the manuscript. RC spectra taken at 6 K show the mismatch between the B excitonic level of the two materials due to the change in the dielectric environment. WSe$_2$ and MoS$_2$, A and B absorption peaks are marked with the dotted lines. This mismatch in the B excitonic level results in an one way ET process from the MoS$_2$-to-WSe$_2$ layer, as observed by a previous report.[8] As a result, MoS$_2$ PL emission decreases in the HS area (pointed by black arrows), but the WSe$_2$ emission is enhanced in the HS region (pointed by white arrows).

The PLE measurements were performed in the range of ~2.12-2.5 eV excitation energies with an average power ~ 40 µW. Figure S5 shows the optical image, RC spectra and PLE maps of the sample.

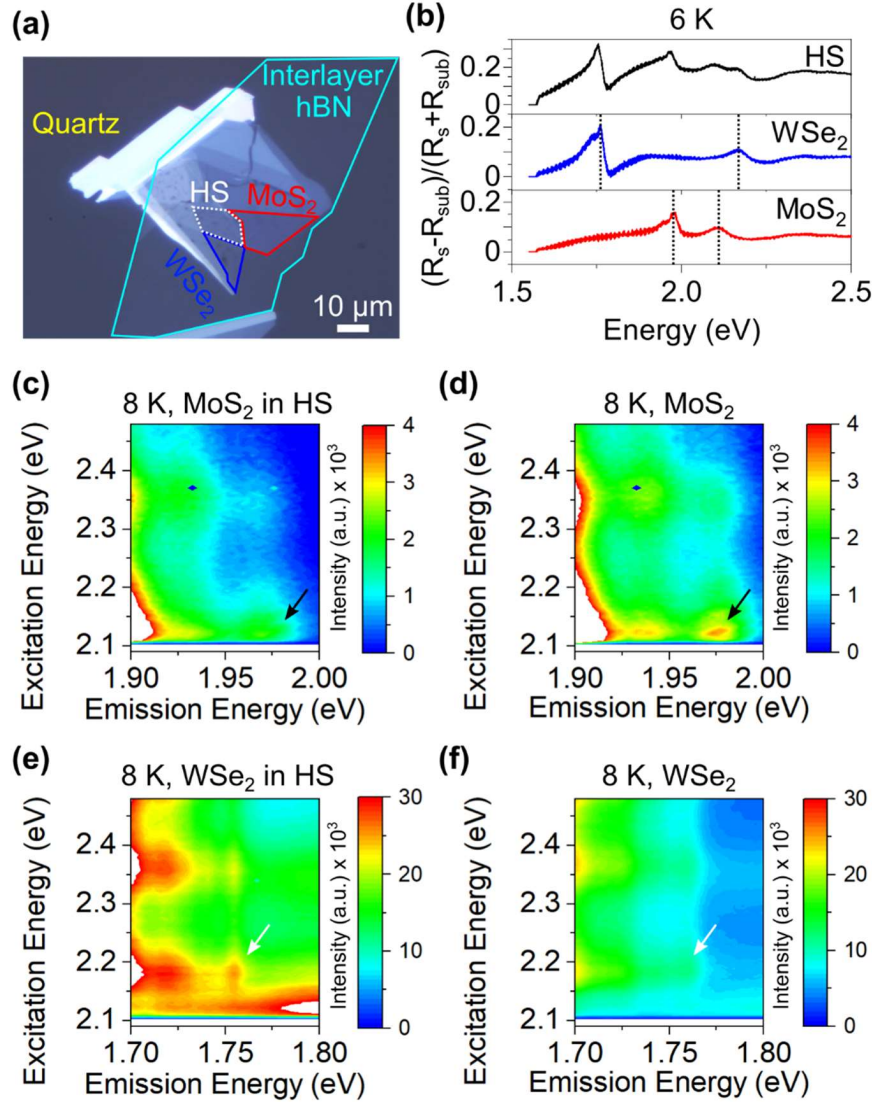

*Figure S5: (a) Optical micrograph of the $MoS_2$-hBN-$WSe_2$ HS on transparent quartz substrate. (b) RC spectra of the sample from the three different regions taken at 6 K. B excitonic resonance between the two material breaks due to the change in dielectric environment. (c)-(d) PLE maps of the $MoS_2$ in HS and 1L $MoS_2$ taken at 8 K, respectively. $MoS_2$ PL emission does not show an increased intensity in the HS area. Both the plots have the same intensity range. Black arrows indicate the $MoS_2$ excitonic emission. (e)-(f) PLE maps of the $WSe_2$ in HS and 1L $WSe_2$ taken at 8 K, respectively. $WSe_2$ emission in the HS area shows an enhanced intensity, proving that only one-way ET happened from the $MoS_2$-to-$WSe_2$ layer. Both the maps have the same intensity range. White arrows indicate the $WSe_2$ excitonic emission.*

**PLE maps of the HS and MoS$_2$ area taken at 200 k from the main sample (Figure S6):**

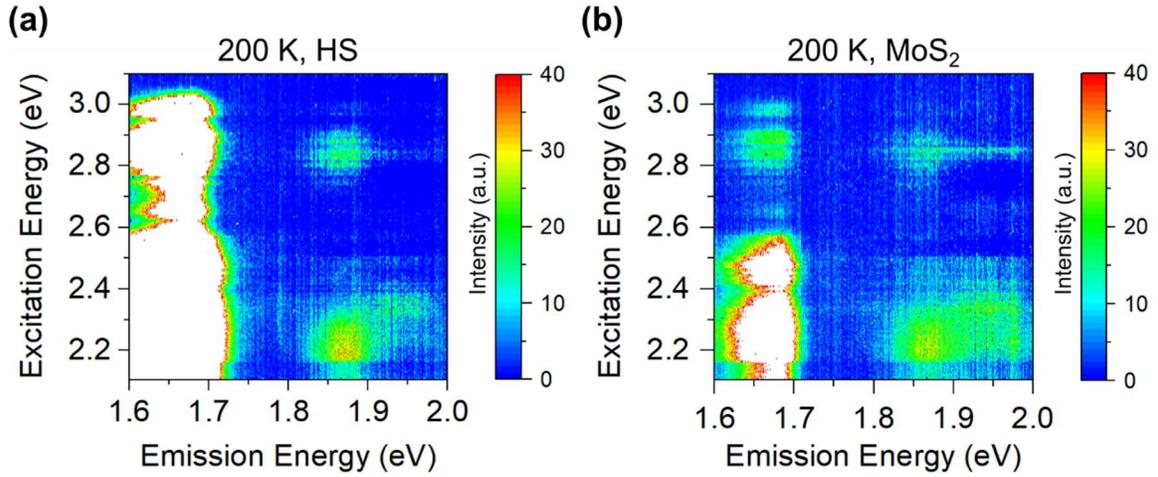

*Figure S6: (a)-(b) PLE maps of the HS and MoS$_2$ at 200 K, respectively. MoS$_2$ PL emission does not increase in the HS area. WSe$_2$ emission in the HS data is saturated to visualize the MoS$_2$ emission. Both the plots have the same intensity range.*

**HS and MoS$_2$ PL intensity comparisons at WSe$_2$ D and B excitations, respectively (Figure S7):**

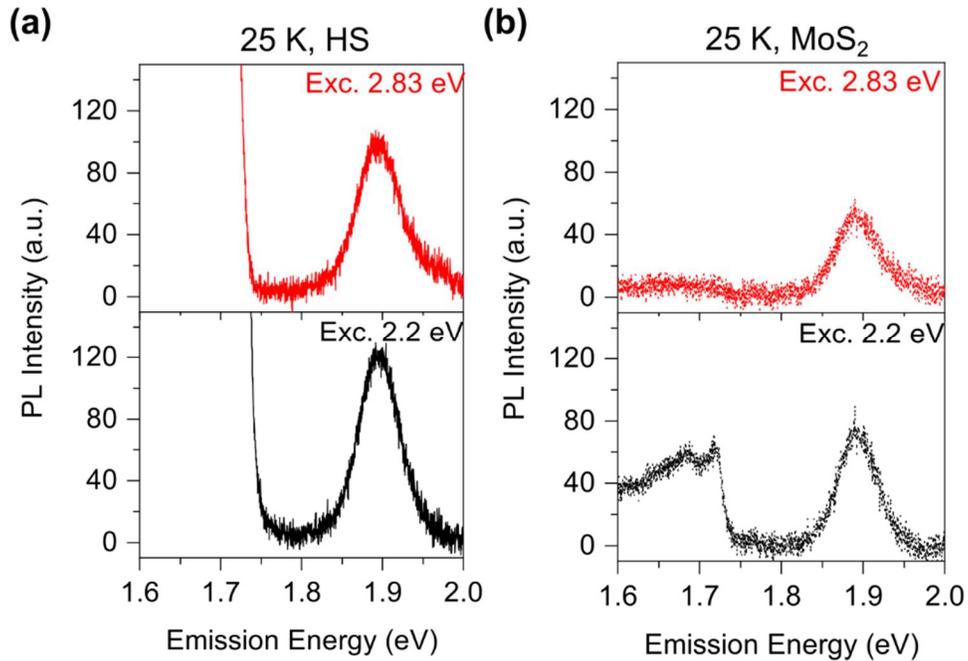

*Figure S7: (a) Top and bottom panel shows PL emission of the MoS$_2$ in the HS area under excitation at 2.83 eV and 2.2 eV, respectively. (b) PL emission profile from the 1L MoS$_2$ area under same excitation conditions. MoS$_2$ PL emission in the HS area shows similar enhancement factor of ~ 1.6 at both excitation energies.*

**DFT calculated MoS₂ and WSe₂ spin-resolved energy landscape along the Γ-K direction (Figure S8):**

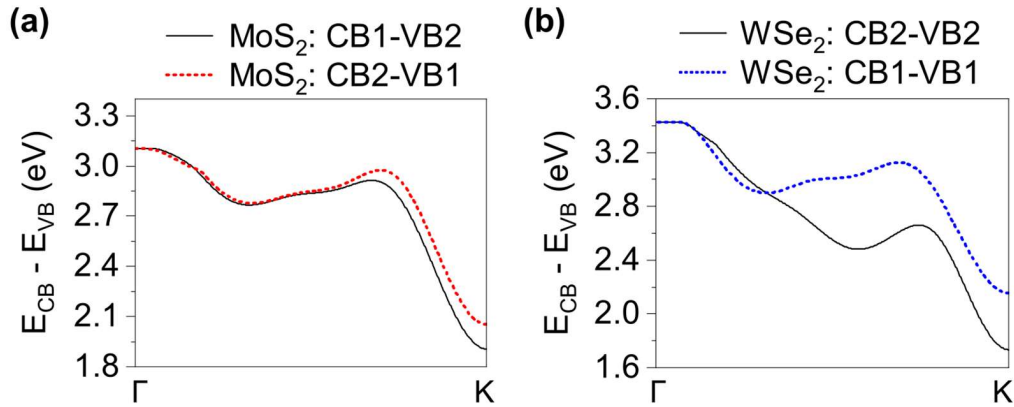

*Figure S8: Calculated spin-resolved momentum-space optical absorption energy landscape of 1L (a) MoS₂ and (b) WSe₂ along the Γ-K direction in the Brillouin zone.*

**PL intensity map of the MoS₂ emission at WSe₂ C excitation (Figure S9):**

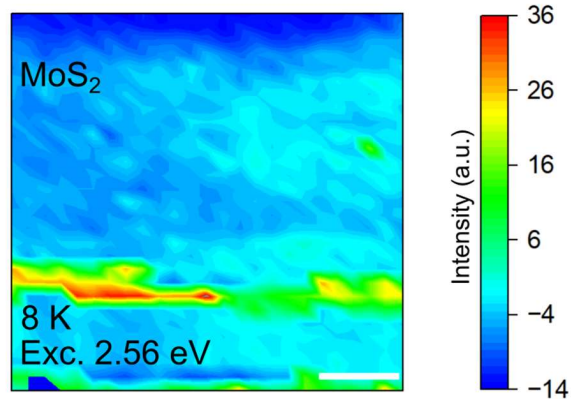

*Figure S9: PL Intensity maps at the resonant WSe₂ C excitation (~ 2.56 eV). MoS₂ does not show any PL emission at this excitation energy. Only system noise was detected in this condition. Scale bar represents 5 μm length.*

**References:**

(1) Clark, S. J.; Segall, M. D.; Pickard, C. J.; Hasnip, P. J.; Probert, M. I. J.; Refson, K.; Payne, M. C. First Principles Methods Using CASTEP. *Zeitschrift für Kristallographie - Crystalline Materials* **2005**, *220* (5–6), 567–570. https://doi.org/10.1524/zkri.220.5.567.65075.

(2) Perdew, J. P.; Burke, K.; Ernzerhof, M. Generalized Gradient Approximation Made Simple. *Phys. Rev. Lett.* **1996**, *77* (18), 3865–3868. https://doi.org/10.1103/PhysRevLett.77.3865.

(3) Tkatchenko, A.; Scheffler, M. Accurate Molecular Van Der Waals Interactions from Ground-State Electron Density and Free-Atom Reference Data. *Phys. Rev. Lett.* **2009**, *102* (7), 073005. https://doi.org/10.1103/PhysRevLett.102.073005.

(4) McNellis, E. R.; Meyer, J.; Reuter, K. Azobenzene at Coinage Metal Surfaces: Role of Dispersive van Der Waals Interactions. *Phys. Rev. B* **2009**, *80* (20), 205414. https://doi.org/10.1103/PhysRevB.80.205414.

(5) Delley, B. Hardness Conserving Semilocal Pseudopotentials. *Phys. Rev. B* **2002**, *66* (15), 155125. https://doi.org/10.1103/PhysRevB.66.155125.

(6) Lin, J. S.; Qteish, A.; Payne, M. C.; Heine, V. Optimized and Transferable Nonlocal Separable Ab Initio Pseudopotentials. *Phys. Rev. B* **1993**, *47* (8), 4174–4180. https://doi.org/10.1103/PhysRevB.47.4174.

(7) Corso, A. D.; Conte, A. M. Spin-Orbit Coupling with Ultrasoft Pseudopotentials: Application to Au and Pt. *Phys. Rev. B* **2005**, *71* (11), 115106. https://doi.org/10.1103/PhysRevB.71.115106.

(8) Kozawa, D.; Carvalho, A.; Verzhbitskiy, I.; Giustiniano, F.; Miyauchi, Y.; Mouri, S.; Castro Neto, A. H.; Matsuda, K.; Eda, G. Evidence for Fast Interlayer Energy Transfer in MoSe2/WS2 Heterostructures. *Nano Lett.* **2016**, *16* (7), 4087–4093. https://doi.org/10.1021/acs.nanolett.6b00801.